\documentstyle[12pt,psfig]{article}
\newcommand{\nc}{\newcommand}
\nc{\beq}{\begin{equation}} \nc{\eeq}{\end{equation}}
\nc{\beqa}{\begin{eqnarray}} \nc{\eeqa}{\end{eqnarray}}
\nc{\lsim}{\begin{array}{c}\,\sim\vspace{-21pt}\\< \end{array}}
\nc{\gsim}{\begin{array}{c}\sim\vspace{-21pt}\\> \end{array}}
\newcommand{\drawsquare}[2]{\hbox{%
\rule{#2pt}{#1pt}\hskip-#2pt%  left vertical
\rule{#1pt}{#2pt}\hskip-#1pt%  lower horizontal
\rule[#1pt]{#1pt}{#2pt}}\rule[#1pt]{#2pt}{#2pt}\hskip-#2pt
\rule{#2pt}{#1pt}}% right vertical
\newcommand{\Yfund}{\raisebox{-.5pt}{\drawsquare{6.5}{0.4}}}
\begin{document}

\begin{titlepage}

\begin{center}

\vspace{2cm}

{\hbox to\hsize{hep-th/9608157 \hfill EFI-96-32}}
{\hbox to\hsize{August 1996 \hfill MIT-CTP-2554}}

\vspace{2cm}

\bigskip

{\Large \bf   Holomorphic Anomalies and the Nonrenormalization
Theorem}

\bigskip

\bigskip

{\bf Erich Poppitz}$^{\bf a}$\footnote{Robert R. McCormick Fellow. 
Supported in 
part by DOE contract DF-FGP2-90ER40560.}  
and  {\bf Lisa Randall}$^{\bf b}$\footnote{NSF Young Investigator 
Award, 
Alfred P. Sloan
Foundation Fellowship, DOE Outstanding Junior Investigator 
Award.}\\

\bigskip

$^{\bf a}${\small \it Enrico Fermi Institute\\
 University of Chicago\\
 5640 S. Ellis Avenue\\
 Chicago, IL 60637, USA\\}

{\tt epoppitz@yukawa.uchicago.edu\\}
\smallskip

 \bigskip

$^{\bf b}${ \small \it   Center for Theoretical Physics\\
    Laboratory of Nuclear Science and Department of Physics\\
    Massachusetts Institute of Technology\\
Cambridge, MA 02139, USA \\}

  {\tt  lisa@ctptop.mit.edu } 

 \bigskip

 \bigskip

\vspace{2cm}

{\bf Abstract}

\end{center} 

It has been argued that the superpotential
can be renormalized in the presence of massless particles.
Possible implications  which have been considered
include the restoration of supersymmetry at higher loops or 
a shift to a supersymmetric vacuum state.  
We argue that even in the presence of massless particles,
there are no new contributions to the superpotential  
at any order in perturbation theory. 
This confirms the utility of the Wilsonian superpotential 
for analyzing the moduli space of the low energy theory.

 \end{titlepage}

\renewcommand{\thepage}{\arabic{page}}

\setcounter{page}{1}

\setcounter{equation}{0}

\baselineskip=18pt

The supersymmetric nonrenormalization theorem \cite{GRS} 
states 
that all perturbative
corrections to the effective action come as contributions to 
the effective K\" ahler
potential. In the presence of  massless particles, however, it 
was noted 
in refs.~\cite{SV}-\cite{W2}, that certain infrared divergent 
contributions 
to the K\" ahler
potential appear to be effectively superpotential renormalizations.
An infrared divergent contribution
to the K\" ahler potential  of the following form
can be rewritten as a superpotential interaction
\beq
\label{deltaw}
\int d^4 \theta ~{D^2 \over \Yfund } ~ g(\Phi) ~ \sim ~ 
\int d^2 \theta ~g(\Phi)~,
\eeq
by using $[D^2, \bar{D}^2] \sim \Yfund$  and integrating by 
parts in 
superspace \cite{WB} 
($D$ is the supercovariant derivative).  

Some of the 
infrared divergent graphs that generate terms 
of the form (\ref{deltaw}) were
explicitly calculated in refs.~\cite{W1}-\cite{W2}.  The simplest 
example is  the massless Wess-Zumino model with superpotential 
$W = \lambda \Phi^3$. With only $\langle \Phi \Phi^\dagger \rangle$ 
propagators and chiral or antichiral vertices, the simplest graph
that has only external $\Phi$ lines arises at two loops (fig.~1). 
The calculation of ref.~\cite{JJW}  
shows that it contributes an expression 
like eq.~(\ref{deltaw}),  with 
$g = \lambda (\lambda^\dagger \lambda)^2 \Phi^3$, 
which renormalizes the superpotential and has a nonholomorphic 
dependence on the couplings.

The nonrenormalization theorem of refs.~\cite{SV}, \cite{Seiberg}, 
which 
forbids such nonholomorphic dependence, 
holds for the Wilsonian effective action, where the infrared 
effects 
are  cut off, and consequently terms like (\ref{deltaw}) 
cannot appear. 
In various applications, however, when one 
is interested in finding the ground state of the theory, the 
object  to minimize is the one-particle-irreducible (1PI) 
effective action, 
which is obtained from the Wilsonian effective action
 after integrating over all momenta, including the infrared. 
In the 1PI effective action, terms like (\ref{deltaw}) are 
allowed, 
and renormalization of the superpotential is possible.
In theories with no massless particles, the superpotentials 
of the Wilsonian and 1PI effective actions coincide, and 
use of the 
Wilsonian superpotential for finding the (supersymmetric)
ground state is thus justified. In theories with massless 
particles, however, 
a perturbative renormalization of the superpotential, 
should it exist, would 
contribute to the scalar potential. In particular, the 
renormalization of the 
superpotential, eq.~(\ref{deltaw}),  could in principle  
generate new terms that are 
consistent with the symmetries of the tree level Lagrangian, 
even if they
are omitted from the bare action. 

Such a renormalization of the  superpotential, if it occurred 
for arbitrary 
values of the fields, would have  important consequences for 
supersymmetric
model building, e.g. for supersymmetric theories of flavor, 
or for models  of 
supersymmetry breaking, since the vacuum
could change when higher loops are included.  For example, 
many recent theories 
of dynamical 
supersymmetry breaking employ strong gauge dynamics to break 
supersymmetry \cite{models}. Below the strong coupling scale 
of the 
gauge theory, the dynamics are typically described in terms 
of effective, 
nonrenormalizable, multi-field, O'Raifeartaigh-type models. 
{}From  
the inconsistency of the $F$-term equations of motion in
these effective theories, one concludes that the models 
break supersymmetry. 
The implicit assumption in  this analysis is that the 
superpotentials of the 
Wilsonian and 1PI effective actions  coincide. If they do, 
then the conclusion 
that supersymmetry is broken will hold in the 1PI effective 
action as well. 
If, on the other hand, the superpotential in the 1PI action 
is renormalized, for 
arbitrary field values, there exists the possibility that 
supersymmetry  
is restored after the infrared effects are taken into account, 
as suggested 
in ref.~\cite{DJJ}. We note that a renormalization of the
superpotential of the type discussed above
  would also contribute to a shift of the vacuum in the 
case of unbroken supersymmetry.

\begin{figure}
\psfig{file=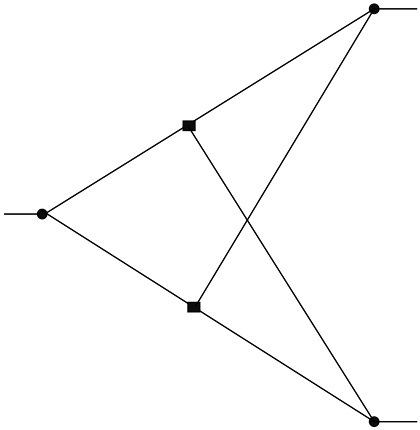}
\caption{The lowest order graph with only chiral external 
lines in the massless 
Wess-Zumino model that  generates a term of the form (\ref{deltaw}), with 
$g(\Phi) = \lambda (\lambda^\dagger \lambda)^2 \Phi^3$.
 The circles and squares 
denote $\lambda \Phi^3$ and  $\lambda^\dagger \Phi^{\dagger 3}$ 
vertices, 
respectively.}
\end{figure}
To elucidate this possibility, consider
the general form of the 1PI effective potential:
\beq
\label{generalveff}
V_{eff} = -  F^* F ~ \left[ f_1(A^*, A)  + f_2(A^*, A, F^*, F) 
\right] - 
\left[ F^* ~ w^{\prime *} (A^*) +  F^* ~ g^{\prime *} (A^*) + 
{\rm h.c.} \right] ~,
\eeq
where $A, F$ denote the scalar and auxiliary components of the 
chiral superfields in the 
model.  The function $f_1$ is the K\" ahler contribution to the 
effective potential, while 
 $f_2$ is the non-K\" ahlerian contribution (i.e. the one that 
 depends on the 
chiral superfields as well as on their supercovariant derivatives); 
some non-K\" ahlerian contributions to the effective potential 
have
been calculated in the framework of the 
(supercovariant) derivative expansion, see e.g. \cite{PW} and 
references therein. 
Finally, $w^\prime$ denotes the gradient
of the tree-level superpotential, and $g^\prime$ is the 
gradient of  a possible 
renormalization of the superpotential
due to infrared divergent contributions like the one in 
eq.~(\ref{deltaw}). {}From eq.~(\ref{generalveff}) one concludes that 
if $F = 0 $ is not a solution of the tree level equations of 
motion,
and the superpotential renormalization is absent ($g = 0$),  
$F=0$ cannot be a minimum of the full 1PI effective potential 
as well, 
and thus supersymmetry can not be restored quantum-mechanically. 
On the other hand, if there is superpotential  renormalization, 
$g\ne 0$, which does not vanish at field values for which 
$g^\prime \ne 0$, 
it is in principle possible to have an $F=0$ minimum of the 
 1PI effective 
potential (\ref{generalveff}),  and hence supersymmetry 
restoration. 

Since we are interested in the question of supersymmetry
restoration, the $F$ terms can be nonzero.  In general,
the supersymmetry breaking $F$ terms can be thought of
as spurions, and one can sum over all graphs with such
terms inserted. Alternatively, one can work in a background
source which restores supersymmetry, and all nonsupersymmetric
diagrams would be proportional to powers of the source.
In general, the result of a loop diagram is quadratic in $F$.
Note that in the presence of supersymmetry breaking,
there can be logarithmic dependence on $F$, but
the overall factor of $|F|^2$ means that these are
not relevant to supersymmetry restoration (or shifting the
supersymmetric vacuum). The possible exception is if there are 
additional holomorphic anomalies
 proportional to $1/F$, 
arising from $D$-terms 
such as 
\beq
\label{deltaw2}
\int d^4 \theta~ {\Phi^\dagger \over \bar{D}^2 \Phi^\dagger}~
 g(\Phi) ~\sim \int d^2 \theta ~g(\Phi)~.
\eeq 
Although  we
know of no such example in the literature,
we also consider this possibility. The dimensional argument we
give below is readily generalized to show that no 
holomorphic anomalies of this sort occur.

A superpotential renormalization, 
if it existed for field values such that $g^\prime(A) \ne 0$, 
could  also cause a shift in a supersymmetric vacuum.  Our argument
also implies the vacuum is not shifted by infrared divergent loop effects 
 in the supersymmetric case.

In our subsequent discussion we focus on the effective action in a 
supersymmetry preserving (i.e. $\theta$-independent) background. 
We note that even in theories 
with broken tree level supersymmetry such an expansion 
is possible, since one can always adjust the source terms 
so that the $F$-components of all background fields vanish. 
The effective action will then have a complicated dependence on 
the sources. However, we are only interested in linear terms in 
the sources (or, equivalently, in the limit of vanishing sources, terms linear
 in the $F$ components of the background),
since, as discussed above, they are the ones that are relevant for
supersymmetry restoration, or for  a shift of the supersymmetric vacuum.
These terms can be obtained by keeping the linear term in the 
(supercovariant) derivative expansion; when a source
is added to keep the $F$ components of the background vanishing,
this is equivalent to keeping the linear terms in the sources. 
We note that full dependence of the effective action
on the $F$ terms of the background field can
in principle be obtained by summing
up all orders in the sources (or equivalently in the derivative expansion). 

We now consider the possible renormalizations to the 
superpotential.
We will argue that  when one properly   analyzes the infrared
structure of the theory using the observations of  Coleman 
and Weinberg
\cite{colemanweinberg}, these renormalizations do not occur. 
Although the 
fields are massless at vanishing field value,
nonzero  field values   induce masses which effectively
cut off   infrared divergent loops. When all the graphs are 
appropriately
summed, there are no new contributions to the superpotential 
that can 
take nonzero value.

We consider a general effective (non-renormalizable) 
Wess-Zumino type theory
with chiral superfields (for brevity of notation hereafter 
$\{ \Phi \}$ denotes a 
generic superfield).
In the background field expansion the renormalization
of the superpotential 
(\ref{deltaw}) would arise from terms of the form:
\beq
\label{deltaw1}
\int d^4 \theta ~{D^2 \over \Yfund +  |W''(\Phi)|^2 } ~ 
g(\{ \Phi \} ) ~.
\eeq
It is clear that for general values of $|W''(\Phi)|$, there
are no infrared divergences (eq.~(\ref{deltaw1})  is schematic 
and is only intended to
illustrate our point: in a general background all propagators 
would be massive, 
proportional to $( \Yfund +  |W''(\Phi)|^2 )^{-1}$, and the 
contribution of a graph 
like the one from fig.~1 could not in general be rewritten in 
this simple form). 
Such a term would not renormalize
the superpotential (for example, for nonvanishing $W''(\Phi)$, 
in the zero-momentum limit, 
eq.~(\ref{deltaw1}) can be seen to contribute to the
non-K\" ahlerian term). 
Since a term in the superpotential should
be present independent of the value of the fields, one
would conclude that there are no renormalizations
of the superpotential; this operator can only be interpreted
as a $D$-term (non-K\" ahlerian) renormalization. 
However, the physically relevant issue is
whether there is a particular field value for
which this term is responsible for supersymmetry
restoration or a shift of the supersymmetric vacuum.
We will now argue that this never occurs. In fact,
we will  show  that the only place where holomorphic
anomalies could have been relevant  to determining the vacuum (that is, where
the field-dependent mass vanishes) is precisely   where
the operator which would have been generated gives
no contribution to the $F$-term of any field (that is, a 
sufficient number
of external fields vanish).    
 
The main idea behind the argument is that in computing
the 1PI effective potential \cite{colemanweinberg}, 
one sums graphs with an arbitrary number of insertions
of external lines, which provide an effective infrared cutoff on 
the propagators. 
When one turns
on an arbitrary background field, 
all fields that couple nontrivially, and 
could  therefore appear in a Feynman graph, obtain mass;
hence in an arbitrary background there cannot be a holomorphic 
anomaly of the 
type  (\ref{deltaw}). Terms like eq.~(\ref{deltaw}) 
can only be generated for specific values of the background
fields, e.g.  when $W''(\Phi)=0$ in eq.~(\ref{deltaw1}). The 
real physical issue is whether there exist field values 
for which the contribution to $g^\prime$ in eq.~(\ref{generalveff}) would 
be nonvanishing, so that these terms are relevant for determining the vacuum of the theory 
(i.e. for supersymmetry restoration or
 a shift of the supersymmetric vacuum, as
discussed above).
We will argue below, that at these
special field values the superpotential renormalization 
does not contribute to determining the vacuum of the theory. 

This is easily seen for  the example of the Wess-Zumino model, 
considered above. In order for the 
superpotential renormalization, 
$\delta W \sim \lambda (\lambda^\dagger \lambda)^2 \Phi^3$, 
to contribute to the vacuum 
 energy, the $F$ term of the field $\Phi$ has to be nonzero; 
hence the expectation value of $\Phi$ has to be nonvanishing. 
But for $\Phi \ne 0$ the propagator in 
the nonvanishing $\Phi$ background is massive; there is no 
infrared divergence
and terms like (\ref{deltaw}) are consequently not generated. 

It is straightforward to generalize this argument  to 
a model with  vertices of arbitrary dimension  
and arbitrary field content. 
Consider first the contribution of graphs with internal massless 
lines only.
We wish to  generate a term of the 
form (\ref{deltaw}), where $g(\Phi )$ has dimension 3, 
which can contribute to the vacuum energy, so that it is relevant, 
for example, for supersymmetry restoration.
If two or more of the fields in $g(\Phi)$ have
vanishing expectation values, 
the superpotential  term\footnote{It is clear that, 
in the framework of the loop expansion, $g(\Phi)$ 
is polynomial in the fields: it can not be generated at one loop, 
and at each higher
loop order there is only a finite number of prototype graphs
\cite{colemanweinberg}, e.g. the one from fig.~1,
 that generate terms like (\ref{deltaw}, 
\ref{deltaw1}).} 
(\ref{deltaw}) 
does not contribute to the vacuum energy,
since the contribution of $g(\Phi)$ to the  $F$-terms of all 
fields  
vanishes (unless some fields are allowed to escape to infinity; 
we only consider theories without classical flat directions 
where runaway 
behavior is not expected to occur).
Therefore, we can allow at most one of the 
fields in $g(\Phi)$ to 
have a vanishing expectation value. 

Consider first a graph with only chiral or antichiral vertices. 
Let the vertices that involve some external lines be of 
the form:
\beq
\label{vertex1}
{\{ \Phi \}^k \over M^{k - 3}}~.
\eeq
Note that if $k - 2$ of the fields $\Phi$ have nonzero 
expectation value, this term 
gives mass  to the other $2$ fields.
Therefore, 
in order for the  vertex (\ref{vertex1}) not to give mass 
to any of the internal lines
when the external fields have expectation values,
we can take as external 
at most $k - 3$ of the fields $\Phi$, if all of them have 
nonvanishing expectation 
values, or $k-2$ fields, if one of them has a zero expectation 
value.\footnote{If the 
mass of a given field $\phi$ arises from a linear combination 
of, say, two 
vertices,  it is possible that the mass could vanish for 
a specific choice 
of {\it nonzero} vacuum expectation values of the background 
fields. However, our power
counting applies in this case as well: the sum of the 
graphs with two internal 
$\phi$ lines, with the two vertices  interchanged,  
is proportional to the mass and 
vanishes.} Thus
the field dependence for the operator of maximum
dimension, which could contribute  to the energy through 
a holomorphic
anomaly, would have the form:
\beq
\label{graph1}
\left( \Pi_i  { \{ \Phi \}^{k_i - 3} \over M^{k_i - 3} } 
\right) ~
 { \{ \Phi \}^{l - 2} \over M^{l - 3} }~,
\eeq
where we allowed at most one external line 
(in the last vertex) 
to have zero expectation value. We now note that the 
term   (\ref{graph1}) has 
dimension 1,  and since there are no mass scales anywhere in 
the graph, the renormalization of the superpotential 
(\ref{deltaw}), which could 
contribute to the energy, vanishes. We note that terms 
of the form (\ref{deltaw}) 
can  be generated at specific field values, but do not 
contribute to 
the vacuum energy.

Above we only considered purely chiral or antichiral vertices. An 
effective theory will also have $D$-type vertices of the form:
\beq
\label{kahlervertex}
{\Phi^{\dagger n+1} \Phi^{m+1} \over M^{n + m}}~.
\eeq
Taking $k$ of the $\Phi$ fields to be external, we 
find that the contribution of this
vertex to the operator (\ref{graph1}) goes like
\beq
\label{kahlerv2}
{\Phi^k \over M^{n + m}}~.
\eeq
The dimension of the operator (\ref{kahlerv2}) is 
$(k - m) - n \le 1 - n  \le 0$, since
if $k - m  = 1$ one needs $n \ge 1$ in order to have at 
least two internal lines (so that 
the graph is 1PI).  
We thus  conclude that insertions of $D$-type vertices 
cannot increase the dimension 
of the operator (\ref{graph1}). Similar arguments show 
that $D$-type
vertices involving supercovariant derivatives can not 
increase the dimension.

We conclude that graphs with only internal massless 
lines do not generate superpotential
renormalizations of the form (\ref{deltaw}) that 
contribute to the energy. Therefore the
nonholomorphic contributions to the superpotential 
generated by graphs with
massless internal lines only can not lead to supersymmetry 
restoration or a shift of a supersymmetric vacuum.

To include possible contributions of internal massive lines 
as well, we note that one 
could potentially increase the dimension of the operator 
(\ref{graph1}) if some of 
the internal lines, which couple to  external vertices, 
are allowed 
to be massive.  We do not see how including heavy states
 can provide the missing dimensional factor,
based on the following nonrigorous reasoning.

We first note that 
these field dependent masses would 
lead
to non-holomorphic field dependence: since the superfield 
propagators
\footnote{The  
$\langle \Phi \Phi \rangle$-propagator at zero momentum, 
which goes like $1/m$,
is not relevant here since in the internal lines of the 
1PI graph momenta are 
nonvanishing, and  a nonholomorphic  dependence on $m$, and 
hence on the fields, 
would be introduced even by purely chiral propagators.}
are proportional to  $(p^2 - m^\dagger m)^{-1}$, the resulting 
expression would depend on $\Phi^\dagger$ as well as $\Phi$ 
and could not be a term in the superpotential.  This is true
if we are trying to make up for the missing positive dimensional
factors. It is possible that there is nonholomorphic field
dependence in the denominator which is compensated
by nonholomorphic field dependence from vertices.
In this case, one can integrate out the massive lines 
(contract them to a point) along the lines we
discuss below for a field independent mass. Then we can work in the 
effective theory, with $D$- and $F$-type vertices generated by 
integrating out the massive fields. In this effective theory, our power
counting arguments above will apply.
 
The only remaining possibility therefore is that  
some of the internal lines  have field independent
masses (due to mass terms in the superpotential) and 
supply the missing dimensionful factor (a larger than 1 
power of the mass) in (\ref{graph1}). 
While we can not rigorously show that this does not occur, 
the following arguments suggest that it is unlikely:
     
In order to cast  (\ref{graph1}) in the form  (\ref{deltaw}),
one needs a factor  of 
\beq
\label{factor}
{m^2 D^2\over \Yfund} ~.
\eeq
However, we believe that  (\ref{factor}) 
cannot be generated by massive propagators in 
the loops. 
We note that in order to obtain the contribution 
(\ref{factor}), 
both large and small momenta  in the loops have to be 
relevant. 
In particular, to obtain a positive power of the mass, the 
momentum 
in the loop that contains the massive propagator has to be 
large (if the 
momentum is small, the corresponding contribution 
is suppressed by inverse powers of the mass; an effective 
(local) higher
dimensional operator is generated, for which our previous 
arguments apply). 
The loop[s] with the massive propagator generates an 
effective operator 
of dimension 2 (i.e. a $D$-term). 
The operator, proportional to  a  positive power of the 
mass  that can be produced by such loops is therefore of  very low 
dimension. In order to embed it in an 1PI graph, the operator 
has to have at least two external fields; therefore 
by dimensional arguments the mass in the numerator has to be
compensated either by mass or momentum in the denominator. 
If it is compensated by mass, we obtain an effective 
local operator for which the previous power counting applies.
The only possibility then is that the loop is proportional 
to a factor of $m^2/p^2$. 
In order to argue that this does not happen, recall that 
the singularities of a 
general Feynman diagram (as a function of external 
momenta)  are determined by the solutions of the Landau 
equations. 
These equations imply that singularities in the external 
momenta  can only occur for values of the loop momenta for which  
some of the propagators are on shell, while all other propagators are 
contracted to points---a momentum configuration corresponding to 
 the so-called reduced diagram (obeying  in addition  
the ``physical picture" condition) \cite{colemannorton}. 
Singularities for which all massive propagators in the graph are 
contracted to points cannot  contribute terms like (\ref{factor}) 
(since they are 
far off-shell and are suppressed by either large momentum 
or large mass), 
while singularities that correspond to a solution of the 
Landau equations 
for which some of the massive propagators are on-shell contribute 
threshold factors and logarithms (see e.g. Sect.~6.3 in 
\cite{iz}) and 
are therefore irrelevant for our discussion.  We thus 
conclude that it is 
unlikely that the inclusion of  massive internal lines will 
supply the missing 
dimensionful factor (\ref{factor}) in eq.~(\ref{graph1}).

Finally, we argue, that even with broken supersymmetry (nonvanishing
$F$ terms) our arguments apply. First we note that a factor of $1/F$,
eq.~(\ref{deltaw2}), 
requires vanishing fields  and momenta as well;  otherwise there is no singularity
for zero $F$. One can therefore proceed as above: first contract
the massive lines and consider a diagram with higher dimensional
operator insertions and with massless or massive internal lines
with supersymmetry breaking masses.  Because the only
terms which interest us are those where the net result of evaluating
the Feynman diagram must be proportional to at most one power of $F$,
the same argument as above shows that the net dimension of
the generated operator will be too low  to contribute
a holomorphic anomaly (again, this applies only for nonvanishing
external fields so that the diagram is relevant to determining the vacuum).

We end  with several comments. 

First, we would like to stress the different aspects
one is interested in when studying the role of 
the holomorphic anomaly in the superpotential 
renormalization (in particular its relevance for determining the
vacuum), discussed in this letter,
 and when studying its contribution to 
 the renormalization of the 1PI 
gauge kinetic function beyond one loop \cite{SV}. 
As discussed above, if the superpotential 
renormalization is to be relevant for determining the
vacuum, one is necessarily interested in nonvanishing field
expectation values. However, 
 at expectation values which are relevant,
 the fields are massive and the holomorphic anomaly is absent. 
In contrast,  a calculation at
zero expectation values  is relevant in the 
context of calculating the gauge beta function. 
Hence, when such a calculation is performed, 
the holomorphic anomaly contributes to the higher 
loop renormalization of the 1PI gauge kinetic function. 
Alternatively, a calculation of the gauge beta function can 
be performed at nonvanishing field expectation values (see
\cite{littlemiracles} and references therein),   
when, as in the case discussed in this letter, 
the holomorphic anomaly contribution to the gauge 
kinetic function is absent and the higher loop 
contributions to the gauge coupling beta function 
are due to wave function renormalizations. 

Second, we note that our analysis above assumed global 
supersymmetry. In supergravity,  in general,  both the  
K\" ahler potential  and superpotential determine whether  
the vacuum is supersymmetric (i.e. the condition 
for $F$ type breaking $W^\prime \ne 0$ is replaced by 
$e^{K/2} (W^\prime + K^\prime W/M_{\rm Planck}^2) \ne 0$ \cite{WB}). 
In this case, our analysis is only relevant 
for a nearly flat K\" ahler metric of the ultraviolet theory,  
at small  (i.e. less than  $M_{\rm Planck}$)  field values, 
where supersymmetry of the vacuum is 
determined only by  the superpotential.

Third, we expect our conclusion that the superpotential 
renormalization does not contribute to the energy 
to hold also if the low-energy theory  involves 
weakly coupled gauge fields, in addition to the chiral matter.
The point is the same: if the gauge field
couples, so that it contributes in a Feynman diagram,
it will get a mass for nonvanishing field values. 
At the point in field space where there are superpotential 
terms, generated by the holomorphic anomaly \cite{W2},
 they would not contribute to the vacuum energy 
(or else, in order to contribute to the $F$-terms, a sufficient 
number of  fields would have to have expectation values, 
making the gauge and matter fields massive,
so that there would be no holomorphic anomaly). We expect 
the rest of the dimensional analysis argument to go through.

Finally, we note that the invariance of the Witten index  does not 
preclude the possibility that supersymmetry is restored when
the infrared effects are taken into account (i.e. in the infinite
volume limit). If a theory breaks supersymmetry, 
the index vanishes, however, the converse is not true: 
the vanishing Witten index does 
not imply that supersymmetry is broken 
in the infinite volume theory \cite{witten}.

To summarize, from general considerations at arbitrary field
values, we conclude that the superpotential is not renormalized.
The previous arguments show furthermore that
there are no new contributions to the vacuum
energy due to the holomorphic anomaly, 
because they necessarily vanish at the point where
holomorphic anomalies would have been relevant. We conclude
that it is safe to address questions such as whether
or not supersymmetry is broken by the vacuum using the
Wilsonian superpotential, and that the only superpotential
nonrenormalizations are nonperturbative.

We thank Tim Jones, Elias Kiritsis, Costas Kounnas, Mikhail Shifman  
and Erick Weinberg for conversations, and the Aspen Center for Physics 
and CERN for their hospitality.

\nc{\ib}[3]{ {\em ibid. }{\bf #1} (19#2) #3}
\nc{\np}[3]{ {\em Nucl.\ Phys. }{\bf #1} (19#2) #3}
\nc{\pl}[3]{ {\em Phys.\ Lett. }{\bf #1} (19#2) #3}
\nc{\pr}[3]{ {\em Phys.\ Rev. }{\bf #1} (19#2) #3}
\nc{\prep}[3]{ {\em Phys.\ Rep. }{\bf #1} (19#2) #3}
\nc{\prl}[3]{ {\em Phys.\ Rev.\ Lett. }{\bf #1} (19#2) #3}


\begin{thebibliography}{99}

\bibitem{GRS}M. Grisaru, M. Ro\v cek and W. Siegel, \np{B159}{79}{429}.

\bibitem{SV}M.A. Shifman and A.I. Vainshtein, \np{B277}{86}{456}; 
\np{B359}{91}{571}.

\bibitem{W1}P. West, \pl{B258}{91}{375}.

\bibitem{JJW}I. Jack, D.R.T. Jones and P. West, \pl{B258}{91}{382}.

\bibitem{DJJ}D.C. Dunbar, I. Jack and D.R.T. Jones, \pl{B261}{91}{62}.

\bibitem{W2}P. West, \pl{B261}{91}{396}.

\bibitem{WB}J. Wess and J. Bagger, {\it Supersymmetry and Supergravity}, 
Princeton UP, (1992).


\bibitem{Seiberg}N. Seiberg, hep-ph/9309335, \pl{B318}{93}{469}.


\bibitem{models}
K. Intriligator, N. Seiberg and S. Shenker, hep-ph/9410203, \pl{B342}{95}{152}; 
M. Dine, A. Nelson, Y. Nir and Y. Shirman, hep-ph/9507378, \pr{D53}{96}{2658}; 
E. Poppitz and S.P. Trivedi, hep-th/9507169, \pl{B365}{96}{125};
P. Pouliot, hep-th/9510148, \pl{B367}{96}{151};
P. Pouliot and M. Strassler, hep-th/9602031, \pl{B375}{96}{175};
T. Kawano, hep-th/9602035;
K.-I. Izawa and T. Yanagida, hep-th/9602180;
K. Intriligator and S. Thomas, hep-th/9603158, hep-th/9608046;
C. Cs\' aki, L. Randall and W. Skiba, hep-th/9605108; 
E. Poppitz, Y. Shadmi and S.P. Trivedi, hep-th/9505113, hep-th/9506184;
C. Cs\' aki, L. Randall, W. Skiba and R.G. Leigh, hep-th/9607021; 
C. Cs\' aki, W. Skiba and M. Schmaltz, hep-th/9507210.

\bibitem{PW}A. Pickering and P. West, hep-th/9604147.

\bibitem{colemanweinberg}S. Coleman and E. Weinberg, \pr{D7}{73}{1888}.

\bibitem{colemannorton}S. Coleman and R.E. Norton, {\em Nuovo Cim.} 
{\bf 28} (1965) 5018.

\bibitem{iz}C. Itsykson and J.-B. Zuber, {\it Quantum Field Theory}, 
McGraw-Hill, (1980).

\bibitem{littlemiracles}M.A. Shifman, Talk at SUSY'96; hep-ph/9606281.

\bibitem{witten}E. Witten, \np{B202}{82}{253}.

\end{thebibliography}
\end{document}